\documentclass[aps,prc,twocolumn,showpacs,preprintnumbers,amsmath,amssymb,
epsf,superscriptaddress,groupedaddress,nofootinbib,tightenlines]{revtex4}

\usepackage{graphicx}
\usepackage{amsmath}
\usepackage{bm}

\def\beqn{\begin{eqnarray}}
\def\eeqn{\end{eqnarray}}
\def\barr{\begin{array}}
\def\earr{\end{array}}
\def\btab{\begin{tabular}}
\def\etab{\end{tabular}}
\def\bite{\begin{itemize}}
\def\eite{\end{itemize}}
\def\bcen{\begin{center}}
\def\ecen{\end{center}}

\def\eq{\begin{equation}}
\def\ee{\end{equation}}
\def\nn{\nonumber}

\def\kdagger{K\hspace{-0.3cm}/}
\def\ndagger{n\hspace{-0.2cm}/}

\def\Idagger{{\rm I}\hspace{-0.18cm}/}

\def\keldagger{k\hspace{-0.2cm}/}

\def\q2dagger{q_2\hspace{-0.35cm}/\;}

\begin{document}


\title{Beam normal spin asymmetry in the equivalent photon approximation}
\author{M. Gorchtein}
\affiliation{Genoa University, Department of Physics, 16146 Genoa, Italy}
\affiliation{California Institute of Technology, Pasadena, CA 91125, USA}
\email{gorshtey@caltech.edu}
\begin{abstract}
The two-photon exchange contribution to the single spin asymmetries 
with the spin orientation normal to the reaction plane is discussed for 
elastic electron-proton scattering in Regge regime. 
For this, the equivalent photon approximation is adopted. 
In this case, hadronic part of the two-photon exchange amplitude describes 
real Compton scattering (RCS). The imaginary part of helicity 
conserving RCS amplitudes are related to total photoabsorption cross section. 
The contribution of the photon helicity flipping amplitudes is estimated by 
the two pion exchange in the $t$-channel. We observe the double logarithmic 
enhancement in the electron mass but find it's contribution to the asymmetry 
negligibly small in the forward kinematics. 
These results are in strong disagreement with the existing calculation.
\end{abstract}
\pacs{12.40.Nn, 13.40.Gp, 13.60.Fz, 14.20.Dh}
\maketitle
\section{Introduction}
\label{sec:intro}
Recently, the new polarization transfer data for the electromagnetic form 
factors ratio $G_E/G_M$ \cite{gegm_exp} raised a lot of interest for the 
two photon exchange (TPE) physics in elastic electron proton scattering. These 
new data appeared to be incompatible with the Rosenbluth data 
\cite{rosenbluth}. A possible way to reconcile the two data sets was 
proposed \cite{marcguichon}, which consists in a more precise account on the 
TPE amplitude, the real part of which enters the radiative 
corrections to the cross section (Rosenbluth) and the polarization cross 
section ratio in a different manner. At present, only the IR divergent part 
of the two photon exchange contribution, 
corresponding to one of the exchanged photon beeing soft,
is taken into the experimental analysis \cite{motsai_maximon}. 
Two model calculations exist for the real part of the 
TPE amplitude \cite{melnichuk,suki}, and they qualitatively confirm that 
the exchange of two hard photons may be responsible for this discrepancy. 
In order to extract the electric form factor in the model independent 
way, one has thus to study the general case of Compton scattering with two 
spacelike photons. These two photon contributions are important for the 
electroweak sector, as well.\\
\indent
In view of this interest, parity-conserving
single spin asymmetries 
in elastic $ep$-scattering with the spin orientation normal to 
the reaction plane regain an attention. These observables are directly 
related to the imaginary part of the TPE amplitude and have been an object 
of theoretical studies in the 1960's and 70's \cite{derujula}. 
By analyticity, the real part 
of the TPE amplitude is given by a dispersion integral over its imaginary 
part. Therefore, a good understanding of this class of observables is 
absolutely necessary. Recently, first measurements of the beam normal 
spin asymmetry $B_n$\footnote{In the literature, also the $A_n$ notation 
for beam normal spin asymmetry or vector analyzing power was adopted.} 
have been performed in different kinematics \cite{bn_exp}. \\
\indent
Though small (several to tens ppm), this asymmetry can be measured with a 
precision of fractions of ppm. 
Before implementing different models for the real part, where an additional 
uncertainty comes from the dispersion integral over the imaginary part, one 
should check the level of understanding of the imaginary part of TPE. These 
checks have been done for the existing data. Inclusion of the elastic 
(nucleon) intermediate state only \cite{afanas} 
led to negative asymmetry of several ppm in 
the kinematics of SAMPLE experiment but was not enough to describe the data. 
The description of the beam 
normal spin asymmetry within a phenomenological model which uses the full set 
of the single pion electroproduction \cite{marcbarbara} did not give 
satisfactory description at any of the available kinematics. Especially 
intriguing appears the situation with the SAMPLE data with electron lab energy 
$E_{lab}=200 $ MeV, which is just above the 
pion production threshold where the theoretical input is well understood. 
On the other hand, an EFT calculation without dynamical pions 
\cite{diaconescu} was somewhat surprisingly very successful in describing this 
kinematics for $B_n$. This success suggests that to the given order of 
chiral perturbation theory, the role of 
the dynamical pions for this observable might be not too large. \\
\indent
Though even at low energies the situation with the imaginary part of TPE 
amplitude is by far not clear, an attention has to be paid to high energies, 
as well, since the dispersion integral which would give us its real part, 
should be performed over the full energy range. Due to relative ease of 
measuring $B_n$ within the framework of parity violating electron scattering, 
new data from running and upcoming experiments \cite{bn_proposals} will 
stimulate further theoretical investigations of this new observable. A 
calculation of $B_n$ in hard kinematics regime at high energy and momentum 
transfer was performed recently in the framework of generalized parton 
distributions (GPD's) and resulted in asymmetries of $\sim1.5$ ppm. 
\cite{jamarcguichon}.\\ 
\indent
Since a ppm effect measurement at high momentum transfers is an extremely 
complicated task, we concentrate in this work on the forward kinematics. 
For this kinematics, a calculation exists \cite{afanas_qrcs}, where an observation 
is made that the contribution of the situation where the exchanged 
photons are nearly real and overtake the external electron kinematics is 
enhanced as $\ln^2(-t/m^2)$, with $m$ the electron mass and $t<0$ the 
elastic momentum transfer. The authors of \cite{afanas_qrcs} take the forward VCS 
amplitude and DIS structure functions as an input, and provide an estimation 
of $B_n$ in this kinematics as large as 25-35 ppm. We demonstrate that 
such a treatment is not adequate and leads to an overestimation of the effect 
by an order of magnitude for the JLab energy $E_{lab}=5.77$ GeV. 
In order to provide an estimate of 
the contribution of this kinematics to $B_n$, we use the most general real 
Compton scattering amplitude. We observe that the double logarithmic 
enhancement is not a consequence of taking the forward limit but a pure 
kinematical effect within the loop integral. We demonstrate that this 
treatment of the problem leads to much smaller values for this 
asymmetry than those quoted in \cite{afanas_qrcs}. We provide furthermore a clear 
explanation of this discrepancy. In fact, it is a consequence of the well 
known fact that the forward doubly virtual Compton scattering amplitude has a 
non-analytic behaviour in the real photon point $Q^2=0$, 
with $Q^2$ the virtuality of the incoming and outgoing photons. 
By now, it has been noticed for 
the low energy limit of the real and (doubly) virtual Compton scattering, 
where even the leading terms in the low energy expansion are sensitive to the 
order of taking either the low energy limit $\nu\to0$ or 
$Q^2\to0$ first \cite{reviewdr}. 
Our observation generalizes this non-analytical behaviour from the kinematical 
point $\nu=0,\,t=0$ to the whole forward axis $t=0$.\\
\indent
The article is organized as follows: in Section \ref{sec:el_ampl}, 
we define the kinematics, general $ep$-scattering amplitude and the 
observables of interest; in Section \ref{sec:tpe}, the two photon exchange 
mechanism and the photons kinematics is studied; the equivalent photons or 
quasi real Compton scattering approximation and its implementation for the 
case of $B_n$ is given in Section \ref{sec:qrcs}; we present our results in 
Section \ref{sec:results} which are followed up by a discussion and a summary.
\section{Elastic $ep$-scattering amplitude}
\label{sec:el_ampl}
In this work, we consider elastic electron-proton scattering process 
$e(k)+p(p)\to e(k')+p(p')$ for which we define:
\beqn
P&=&\frac{p+p'}{2}\nn\\
K&=&\frac{k+k'}{2}\nn\\
q&=&k-k'\;=\;p'-p,
\eeqn
and choose the invariants $t=q^2<0$\footnote{In elastic $ep$-scattering, the 
usual notation for the momentum transfer is $Q^2=-q^2$ but we prefer the 
more general notation $t$ to avoid confusion with the incoming and outgoing 
photon virtualities in forward doubly virtual Compton scattering we will 
be concerned in the following.} and $\nu=(P\cdot K)/M$ as the 
independent variables. $M$ denotes the nucleon mass. They are related to the 
Mandelstam variables $s=(p+k)^2$ and $u=(p-k')^2$ through $s-u=4M\nu$ and 
$s+u+t=2M^2$. For convenience, we also introduce the usual polarization 
parameter $\varepsilon$ of the virtual photon, which can be related to the 
invariants $\nu$ and $t$ (beglecting the electron mass $m$):
\beqn
\varepsilon\,=\,\frac{\nu^2-M^2\tau(1+\tau)}{\nu^2+M^2\tau(1+\tau)},
\eeqn
with $\tau=-t/(4M^2)$. Elastic scattering of two spin $1/2$ particles is 
described by six independent amplitudes. Three of them do not flip the 
electron helicity \cite{marcguichon},
\beqn
T_{no\;flip}&=&\frac{e^2}{-t}
\bar{u}(k')\gamma_\mu u(k)\label{f1-3}
\\
&\cdot&
\bar{u}(p')
\left(\tilde{G}_M \gamma^\mu\,-\,
\tilde{F}_2\frac{P^\mu}{M}\,+\,
\tilde{F}_3\frac{\kdagger P^\mu}{M^2}\right)u(p),\nn
\eeqn
while the other three are electron helicity flipping and thus have in 
general the order of the electron mass $m$ \cite{jamarcguichon}:
\beqn
T_{flip}&=&\frac{m}{M}\frac{e^2}{-t}
\left[
\bar{u}(k')u(k)\cdot\bar{u}(p')\left(\tilde{F}_4\,+\,
\tilde{F}_5\frac{\kdagger}{M}\right)u(p)\right.\nn\\
&&\;\;\;\;\;\;\;\;\;\;\;+\;
\tilde{F}_6\bar{u}(k')\gamma_5u(k)\cdot\bar{u}(p')\gamma_5u(p)
\Big]
\label{f4-6}
\eeqn
\indent
In the one-photon exchange (Born) approximation, two of the six amplitudes 
match with the electromagnetic form factors,
\beqn
\tilde{G}_M^{Born}(\nu,t)&=&G_M(t),\nn\\
\tilde{F}_2^{Born}(\nu,t)&=&F_2(t),\nn\\
\tilde{F}_{3,4,5,6}^{Born}(\nu,t)&=&0
\eeqn
where $G_M(t)$ and $F_2(t)$ are the magnetic and Pauli form factors, 
respectively. For further convenience we define also 
$\tilde{G}_E\,=\,\tilde{G}_M-(1+\tau)\tilde{F}_2$ and 
$\tilde{F}_1\,=\,\tilde{G}_M-\tilde{F}_2$ which in the Born approximation 
reduce to electric form factor $G_E$ and Dirac form factor $F_1$, 
respectively. For a beam polarized normal to the scattering plane, one can 
define a single spin asymmetry,
\beqn
B_n\,=\,
\frac{\sigma_\uparrow-\sigma_\downarrow}{\sigma_\uparrow+\sigma_\downarrow},
\eeqn
where $\sigma_\uparrow$ ($\sigma_\downarrow$) denotes the cross sesction for 
an unpolarized target and for an electron beam spin parallel (antiparallel) 
to the normal polarization vector defined as 
\beqn
S_n^\mu\,=\,
\left(0,\frac{[\vec{k}\times\vec{k'}]}{|\vec{k}\times\vec{k'}|}\right),
\eeqn
normalized to $(S\cdot S)=-1$. Similarly, one defines the target normal spin 
asymmetry $A_n$. It has been shown in the early 1970's \cite{derujula} that 
such asymmetries are directly related to the imaginary part of the $T$-matrix.
Since the electromagnetic form factors and the one-photon exchange amplitude
are purely real, $B_n$ obtains its finite contribution to leading order in the 
electromagnetic constant $\alpha_{em}$ from an interference between the Born 
amplitude and the imaginary part of the two-photon exchange amplitude.
In terms of the amplitudes of Eqs.(\ref{f1-3},\ref{f4-6}), the beam normal 
spin asymmetry is given by:
\beqn
B_n&=&-\frac{m}{M}\sqrt{2\varepsilon(1-\varepsilon)}\sqrt{1+\tau}
\left(\tau G_M^2+\varepsilon G_E^2\right)^{-1}\nn\\
&\cdot&
\left\{
\tau G_M {\rm Im}\tilde{F}_3\,+\,G_E {\rm Im}\tilde{F}_4
\,+\,F_1\frac{\nu}{M}{\rm Im}\tilde{F}_5
\right\}.
\label{eq:bn_general}
\eeqn
\indent
For completeness, we also give here the expression of target normal spin 
asymmetry $T_n$\footnote{Also $A_n$ notation for target normal spin asymmetry 
exists in the literature.} in terms of invariant amplitudes:
\beqn
T_n&=&\sqrt{2\varepsilon(1+\varepsilon)}\sqrt{\tau}
\left(\tau G_M^2+\varepsilon G_E^2\right)^{-1}\label{eq:an_general}
\\
&\cdot&
\left\{
(1+\tau)\left[F_1{\rm Im}\tilde{F}_2-F_2{\rm Im}\tilde{F}_1\right]\right.\nn\\
&&\left.\,+\,\left(\frac{2\varepsilon}{1+\varepsilon}G_E-G_M\right)
\frac{\nu}{M}{\rm Im}\tilde{F}_3
\right\}.\nn
\eeqn
\section{Two photon exchange}
\label{sec:tpe}
\begin{figure}[h]
{\includegraphics[height=2cm]{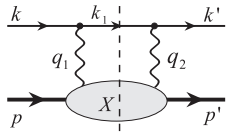}}
\caption{Two-photon exchange diagram.}
\label{boxgraph}
\end{figure}
\begin{figure}[h]
{\includegraphics[height=7cm]{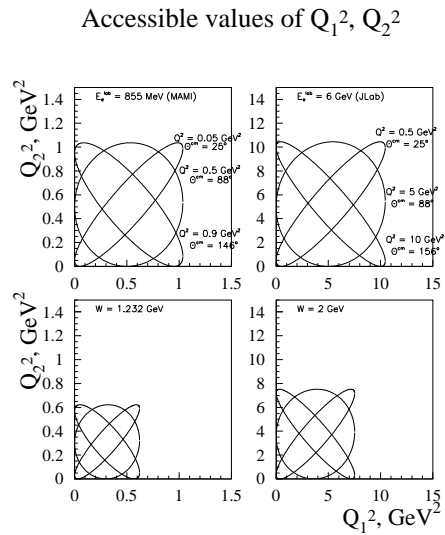}}
\caption{Kinematically allowed values of the photon vitualities $Q_{1,2}^2$.}
\label{fig:kinbounds}
\end{figure}

The imaginary part of the two-photon exchange (TPE) graph in Fig.\ref{boxgraph}
is given by 
\beqn
{\rm Im} {\cal{M}}_{2\gamma}&=&e^2
\int\frac{|\vec{k}_1|^2d|\vec{k}_1|d\Omega_{k_1}}{2E_1(2\pi)^3}
\bar{u}'\gamma_\nu(\keldagger_1+m)\gamma_\mu u\nn\\
&&\;\;\;\;\cdot\,\frac{1}{Q_1^2Q_2^2}W^{\mu\nu}(w^2,Q_1^2,Q_2^2),
\label{eq:im2gamma}
\eeqn
where $W^{\mu\nu}(w^2,Q_1^2,Q_2^2)$ is the imaginary part of doubly virtual 
Compton scattering tensor. $Q_1^2$ and $Q_2^2$ denote the virtualities of 
the exchanged photons in the TPE diagram, and $w$ is the invariant mass of 
the intermediate hadronic system. We next study the kinematics of the 
exchanged photons. Neglecting the small electron mass and using the c.m. frame 
of the electron and proton, one has:
\beqn
Q_{1,2}^2\,=\,2|\vec{k}||\vec{k}_1|(1-\cos\Theta_{1,2}),
\eeqn
with $|\vec{k}|={s-M^2\over2\sqrt{s}}\equiv k$ the three momentum of the 
incoming (and outgoing) eletron, 
$|\vec{k}_1|=\sqrt{({s-w^2+m^2\over2\sqrt{s}})^2-m^2}$ that of the 
intermediate electron, and 
$\cos\Theta_2=\cos\Theta\cos\Theta_1+\sin\Theta\sin\Theta_1\cos\phi$. 
The kinematically allowed values of the virtualities of the exchanged photons 
(the restriction is due to the fact that the intermediate electron is on-shell)
are represented by the internal area of the ellypses shown in 
Fig. \ref{fig:kinbounds}. 
The ellypses are drawn inside a square whose side is 
defined through the external kinematics ($k$) and the invariant mass of the 
intermediate hadronic state ($w^2$ or $k_1$), while the form solely by the 
scattering angle. If choosing higher values of the mass of the hadronic 
system $w^2<s$, it leads to scaling the size of the ellypse by a factor of 
$s-w^2\over s-M^2$. In the limit $w^2=(\sqrt{s}-m_e)^2$, the ellypses shrink 
to a point at the origin and both photons are nearly real. This is not a 
collinear singularity, however, since the real photons' energy remains large 
enough in order to provide the transition from nucleon with the mass $M$ to 
the intermediate state $X$ with the mass $w$. Instead, the intermediate 
electron is soft, $k_1^\mu\approx(m_e,\vec{0})$, therefore this kind of 
kinematics does not lead to an IR divergency which can only occur if the 
intermediate hadronic state is the nucleon itself. In the following we are 
going to study this kinematical situation in more detail.
\section{Quasi-RCS approximation}
\label{sec:qrcs}
We consider the kinematical factors under the integral over the electron phase 
space of Eq.(\ref{eq:im2gamma}) at the upper limit of the integration over the 
invariant mass of the intermediate hadronic state, $w\to\sqrt{s}-m_e$:
\beqn
\frac{\vec{k}_1^2}{E_1}
\frac{1}{Q_1^2Q_2^2}\,\sim\,
\frac{1}{4k^2E_1}
\,\sim\,{1\over m} {1\over4k^2},
\eeqn
In this range of 
the hadronic mass $w$, the exchanged photons are real, and the contribution of 
real Compton scattering (RCS) to the $2\gamma$-exchange graph is enhanced by 
a factor of $1/m$ (it is not a singularity since $B_n$ has a factor of $m$ 
in front). This enhancement, however, only appears for the beam 
asymmetry $B_n$ since the target asymmetry does not involve the electron spin 
flip and the electron mass can be neglected in the electron 
propagator $\keldagger_1+m$ in the numerator of Eq.(\ref{eq:im2gamma}).
The remaining $\keldagger_1$ cancels this $1/m$ behaviour.\\
\indent
We next rewrite the hadronic tensor in Eq.(\ref{eq:im2gamma}) identically as 
\beqn
W^{\mu\nu}(W^2,Q_1^2,Q_2^2)&=&W^{\mu\nu}(s,0,0)\\
&+&
(W^{\mu\nu}(W^2,Q_1^2,Q_2^2)-W^{\mu\nu}(s,0,0))\nn,
\label{eq:rcstensor_reg}
\eeqn
so that only the first term is now enhanced in this limit, while 
the second term vanishes at the RCS point by construction. 
Equivalent photon or quasi-real Compton scattering (QRCS) approximation 
consists in assuming the first term to be dominant due to the kinematical 
enhancement under the integral and in neglecting the second one. The 
question of the validity of such an approximation should be raised for the 
beam normal spin asymmetry. We will come back to this point in Section 
\ref{sec:results}. 
However, in general, this kind of contributions coming from QRCS kinematics 
will allways be present in the full calculation, since the second term in 
Eq.(\ref{eq:rcstensor_reg}) is constructed in such a way that the resulting 
integral is regular at the QRCS point. In the following, the QRCS 
approximation will be used.
Hence, the hadronic tensor can be taken out of the integration over 
the electron phase space. The remaining integrals are
\beqn
I_0&=&\int\frac{d^3\vec{k}_1}{2E_1(2\pi)^3}\frac{1}{Q_1^2Q_2^2}
\,\nn\\
I_1^\mu&=&\int\frac{d^3\vec{k}_1}{2E_1(2\pi)^3}\frac{k_1^\mu}{Q_1^2Q_2^2}
\;=\;K^\mu I_{1K}\,+\,P^\mu I_{1P},
\eeqn
\indent
We next list the result of the integration in the limit of high energies 
(i.e., neglecting terms $\sim m_\pi/E$):
\beqn
I_0&=&
\frac{1}{-32\pi^2t}
\left[
\ln^2\left({-t\over m^2}\right)\,+\,
4{\pi^2\over3}
\right]\;,\nn\\
I_{1P}&=&
\frac{1}{16\pi^2}\frac{s-M^2}{M^4-su}
\ln\left(\frac{4E^2}{-t}\right),
\\
I_{1K}&=&
\frac{1}{-4\pi^2t}
\ln\left(\frac{2E}{m}\right)
\,+\,\frac{4M\nu}{t}I_{1P}\nn\\
&\equiv&I_{1K}^0\,+\,\frac{4M\nu}{t}I_{1P}
\nn.
\eeqn
\indent
For the details, we address the reader to the Appendix.
The three integrals $I_0$, $I_{1K}^0$, and $I_{1P}$ obviously classify as 
$\log^2{m}$, $\log^1{m}$, and $\log^0{m}$, respectively. 
In Ref. \cite{afanas_qrcs} only the first integral was calculated.\\
\indent
The RCS tensor may be taken for instance in the basis of Prange \cite{prange} 
or, equivalently of Berg and Lindner \cite{berglindner},
\beqn
{W}^{\mu\nu}_{RCS}&=&\bar{N}'
\left\{
\frac{P'^\mu P'^\nu}{P'^2}(B_1\,+\,\kdagger B_2)\;+\;
\frac{n^\mu n^\nu}{n^2}(B_3\,+\,\kdagger B_4)\right.\nn\\
&&\;\;\;\;\;\;
\;+\frac{P'^\mu n^\nu\,-\,n^\mu P'^\nu}{P'^2n^2}\,i\gamma_5 B_7\nn\\
&&\;\;\;\;\;\;
\left.\;+\frac{P'^\mu n^\nu\,+\,n^\mu P'^\nu}{P'^2n^2}\,\ndagger B_6
\right\}\,N,
\label{eq:rcstensor}
\eeqn
with the vectors defined as $P'=P-{P\cdot K\over K^2}K$, 
$n^\mu=\varepsilon^{\mu\nu\alpha\beta}P_\nu K_\alpha q_\beta$ such that 
$P'\cdot K = P'\cdot n= n\cdot K=0$. The amplitudes $B_i$ are functions of 
$\nu$ and $t$. This form of Compton tensor is 
convenient due to the simple form of the tensors appearing in 
Eq. (\ref{eq:rcstensor}). However, the amplitudes $B_i$ possess kinematical 
singularities and constraints, therefore in the following we will also use 
another set of invariant amplitudes introduced as combinations of $B_i$'s 
in \cite{lvov}. This latter set of amplitudes is widely used in the 
dispersion analysis of Compton experiments, high energy contributions to 
them are well studied, and we will take an advantage of this knowledge.\\
\indent
Before contracting the leptonic and hadronic tensors,
we notice that the amplitude $B_7$ can only contribute to the invariant 
amplitude $\tilde{F}_6$, since it contains the $\bar{N}'\gamma_5N$ structure.
$\tilde{F}_6$ does not contribute at leading order in $m$
to neither obserevable of interest, therefore $B_7$ will
be neglected in the following. The remaining tensors in Eq.(\ref{eq:rcstensor})
 are symmetric in indices $\mu\nu$.
\beqn
{\rm Im} {\cal{M}}_{2\gamma}^{QRCS}&=&e^2
\bar{u}'\gamma_\nu(\Idagger_1+mI_0)\gamma_\mu u
W_{RCS}^{\mu\nu}\nn\\
&=&
\bar{u}'(-\Idagger_1+mI_0)u
W_{RCS}^{\mu\nu}g_{\mu\nu}\nn\\
&+&2W_{RCS}^{\mu\nu}I_{1\mu}\bar{u}'\gamma_\nu u.
\label{eq:tensor_contraction}
\eeqn
\indent
Finally, we identify 
different terms in Eq. (\ref{eq:tensor_contraction}) 
with the structures, together with amplitudes $\tilde{F}_{1,\dots6}$
parametrizing elastic $ep$-scattering amplitude in Eqs. (\ref{f1-3},\ref{f4-6})
and after some algebra we find for the invariant amplitudes for 
the elastic electron-proton scattering in the QRCS approximation:
\beqn
{\rm Im} \tilde{G}_M^{QRCS}&=&-2tI_{1P}{\rm Im}B_6\label{f1}\\
{\rm Im} \tilde{F}_2^{QRCS}&=&-MtI_{1P}\nn\\
&\cdot&{\rm Im}\left[B_1-B_3-\frac{2Mt}{M^4-su}B_6\right]
\label{f2}\\
{\rm Im} \tilde{F}_3^{QRCS}&=&-M^2tI_{1P}{\rm Im}\nn\\
&\cdot&{\rm Im}\left[B_2-B_4-\frac{8M\nu}{M^4-su}B_6\right],
\label{f3}
\eeqn
\beqn
{\rm Im} \tilde{F}_4^{QRCS}&=&-Mt(I_0-I_{1K}){\rm Im}(B_1+B_3)\nn\\
&-&
4M^2\nu I_{1P}\Big[{\rm Im}(B_1-B_3)\nn\\
&&\left.\;\;\;\;\;\;\;-\,\frac{2Mt}{M^4-su}{\rm Im}B_6\right]
\label{f4}\\
{\rm Im} \tilde{F}_5^{QRCS}&=&-M^2t(I_0-I_{1K}){\rm Im}(B_2+B_4)\nn\\
&-&
4M^3\nu I_{1P}{\rm Im}(B_2-B_4)\nn\\
&-&\frac{2M^2t(4M^2-t)}{M^4-su}{\rm Im}B_6
\label{f5}
\eeqn
\indent
We use next the forward kinematics and obtain for $B_n$ in the QRCS 
approximation:
\beqn
B_n^{QRCS}&=&\frac{m}{M}\frac{\sqrt{-t}}{\nu}
\frac{F_1}{F_1^2+\tau F_2^2}\nn\\
&\cdot&
\left\{
Mt(I_0-I_{1K}){\rm Im}\left[B1+B_3+\nu(B_2+B_4)\right]\right.\nn\\
&&\,+\,\left.
4M^2\nu I_{1P}{\rm Im}\left[B1-B_3+\nu(B_2-B_4)\right]
\right\}\nn\\
&=&-\frac{m}{M}\frac{\sqrt{-t}}{\nu}
\frac{F_1}{F_1^2+\tau F_2^2}\nn\\
&\cdot&
\left\{
Mt^2(I_0-I_{1K}){\rm Im}A_1\right.\nn\\
&&\;\;\;+\,\left.
4M^2\nu I_{1P}4\nu^2{\rm Im}(A_3+A_6)
\right\},
\label{eq:B_n_Ai}
\eeqn
where the invariant amplitudes of Lvov \cite{lvov} are used instead of 
combinations of $B_i$ appearing in the result. We make use of the 
well known physics content of the amplitudes $A_1$ and $A_3+A_6$ entering the 
final result of Eq.(\ref{eq:B_n_Ai}). In terms of the helicity amplitudes of 
real Compton scattering defined as 
$T_{\lambda'_\gamma\lambda'_N,\lambda_\gamma\lambda_N,}\,\equiv\,\varepsilon'^{*\nu}_{\lambda'_\gamma}\varepsilon^\mu_{\lambda'_\gamma}W_{\mu\nu}^{RCS}$, 
one has \cite{lvov}:
\beqn
A_1
&=&\frac{s-M^2}{MQ^2\sqrt{M^4-su}}T_{1{1\over2},-1{1\over2}}
\label{eq:a1_helampl}\\
&+&\frac{1}{2\sqrt{Q^2}(s-M^2)}
\left(T_{-1-{1\over2},1{1\over2}}+T_{1-{1\over2},-1{1\over2}}\right)
,\nn\\
A_3+A_6
&=&\frac{1}{4\nu\sqrt{M^4-su}}
\left(T_{1{1\over2},1{1\over2}}+T_{1{1\over2},1{1\over2}}\right)
,\label{eq:a3a6_helampl}
\eeqn
\indent
It can be seen that the combination $A_3+A_6$
involves only helicity amplitudes without helicity flip, while 
$A_1$ does flip photon helicity. Basing on the 
analysis of RCS \cite{lvov}, we know furthermore that the real part of 
Re$(A_3+A_6)$ is related to the forward nucleon 
polarizability $(\alpha+\beta)$, while Re$A_1$ is 
related to the backward polarizability $(\alpha-\beta)$. It is interesting 
to observe that the backward RCS physics enters the expressions for 
forward $ep$-scattering amplitudes. This result is in fact model independent, 
since the contribution from the QRCS peak is always present in the full result.
The role of the QRCS approximation used here is in neglection of the other 
contributions which might obscure this important observation. Moreover, we 
see in Eq.(\ref{eq:B_n_Ai}) that the forward combination of the RCS 
amplitudes enters $B_n$ multyplied by the 
integral $I_{1P}$ which is peaked in the forward direction, as well, and the 
backward combinations with the backward peaked integrals $I_0-I_{1K}$. 
This result is in disagreement with the result of Ref. \cite{afanas_qrcs}, where 
the $ep$-scattering amplitude in the forward direction is parametrized through
the total photoabsorption cross section and the backward integral $I_0$. 
This discrepancy will be discussed in more detail in Section 
\ref{sec:discussion}.\\
\indent
In the following, we use the total photoabsorption cross section 
as the phenomenological input to Im$(A_3+A_6)$ in the forward 
regime \cite{lvov}
\beqn
{\rm Im}(A_3+A_6)(s,t=0)
&=&-{1\over2\nu}\sigma_{\gamma N}^{tot}
\left({s\over s_0}\right)^{\alpha_P(0)-1}\,.
\eeqn
with the total photoabsorption cross section 
$\sigma_{\gamma N}^{tot}\approx0.1$ mbarn. Furthermore,
we adopted the standard Regge behaviour in order to assure a correct 
energy dependence with the pomeron intercept $\alpha_P(0)=1.08$ and the 
parameter $s_0=1$ GeV. For $t$ dependence, we use an exponential 
suppression factor $\exp(Bt/2)$ with $B=8$ GeV$^{-2}$, which provides a good 
description of 
Compton cross section for $-t\leq0.8$ GeV$^2$ \cite{compton_exp}.\\
\indent
The amplitude $A_1$ in Regge regime is known to have the quantum numbers of a 
scalar isoscalar exchange in the $t$-channel which was successfully described 
by $\sigma$-meson \cite{lvov} or equivalently, by two pion exchange in the 
$t$-channel \cite{jarcs}. Since the effective $\sigma$-meson exchange does not 
result in a non-zero imaginary part in the $s$-channel, one should use the 
two pion exchange mechanism accompanied by multiparticle intermediate state in 
the $s$-channel. In this work we estimate $A_1$ through $\pi N$ and $\rho N$ 
contributions within the ``minimal'' Regge model for $\pi$ and $\rho$ 
photoproduction \cite{marcmichel_piprod} where reggeized description of 
high energy pion 
production is obtained by adding the $t$-channel meson exchange amplitude and 
(in the case of $\gamma p$ reaction) $s$-channel Born amplitude which is 
necessary to ensure gauge invariance. The reggeization procedure naturally 
leads to the substitution of each $t$-channel Feynman propagator by its 
Regge counterpart,
${1\over t-m_\pi^2}\to {\cal P}_\pi^R(\alpha_\pi(t))$, with 
\beqn
{\cal{P}}_R^\pi\,=\,
\left(\frac{s}{s_0}\right)^{\alpha_\pi(t)}
\frac{\pi\alpha'_\pi}{\sin{\pi\alpha_\pi(t)}}
\frac{1}{\Gamma(1+\alpha_\pi(t))}\,,
\eeqn
with $\alpha_\pi(t)=\alpha'_\pi(t-m_\pi^2)$ and $\alpha'_\pi=0.7$ GeV$^{-2}$. 
Gauge invariance requires the $s$-channel piece to be reggeized in the 
same way, i.e., to be multiplied by  
$(t-m_\pi^2){\cal P}_\pi^R(\alpha_\pi(t))$.
Here we list the results of the calculation of $A_1$:
\beqn
{\rm Im}A_1^{\pi N}&=&\frac{2m_\pi^2C_\pi^2}{M(s-M^2)}(B_0^\pi-M^2A_0^\pi),
\label{eq:a1_pi}\\
{\rm Im}A_1^{\rho N}&=&\frac{m_\rho^2C_\rho^2}{2M(s-M^2)}\label{eq:a1_rho}
\left\{{M^2\over s-M^2}C_0^\rho\right.\nn\\
&&+\,\left.{s-3M^2
+5m_\rho^2\over2}B_0^\rho+m_\rho^2{s+M^2\over2}A_0^\rho
\right\},\nn
\eeqn
where $C_\pi=2\sqrt{2}Me{f_{\pi NN}\over m_\pi}$ with $f_{\pi NN}^2/4\pi=0.08$,
and $C_\rho=2\sqrt{2}Me{f_{\pi NN}\over m_\pi}{f_{\rho\pi\gamma}\over m_\pi}$ 
with $f_{\rho\pi\gamma}=0.103$ \cite{marcmichel_piprod}. Where possible, 
$m_\pi$ and $Q^2$ were neglected as compared to $s$, $M$, $m_\rho$ in order to 
simplify the final expression. The scalar integrals 
$A_0$, $B_0$, $C_0$ for both $\pi N$ and $\rho N$ contributions are given in 
the Appendix. 
In Fig. \ref{feynregge}, we show the ratio of the reggeized version of the 
integrals $(A,B,C)^\rho$ and the analytic results. 
\begin{figure}[h]
\vspace{-1cm}
{\includegraphics[height=8cm]{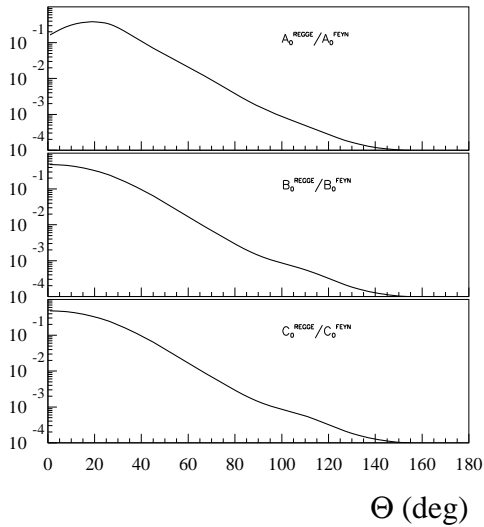}}
\caption{Ratio of the integrals with Regge propagators in the $t$-channel 
to the analytic results with pion Feynman propagators is shown for 
$A_0^\rho$ (upper panel), $B_0^\rho$ (upper panel), $C_0^\rho$ (upper panel) 
as function of c.m. scattering angle for beam energy $E_{lab}=5.77$ GeV.}
\label{feynregge}
\end{figure}
The amplitude $A_1$ is defined as the combination 
${1\over -t}[B_1+B_3+\nu(B_2+B_4)]$, and the singularity ${1\over t}$ in 
the definition only cancels if taking the gauge invariant combination as 
described in \cite{marcmichel_piprod}. The other feature of the result of 
Eqs.(\ref{eq:a1_pi},\ref{eq:a1_rho}) is that both contributions are 
suppressed by factors ${m_\pi^2\over s-M^2}$ and $m_\rho^2\over s-M^2$, 
respectivelyly. In the case of the pion, it is interesting to observe this 
fact in view of somewhat surprising success of the effective field description 
of the SAMPLE data point on $B_n$ without pion contribution. This might give 
a hint that the pion contribution to $B_n$ at low energies is suppressed by 
the pion mass, if calculated to the same order.
\section{Results}
\label{sec:results}
We now present the results for $B_n$ using the model 
described in the previous section.
\begin{figure}[h]
{\includegraphics[height=8cm]{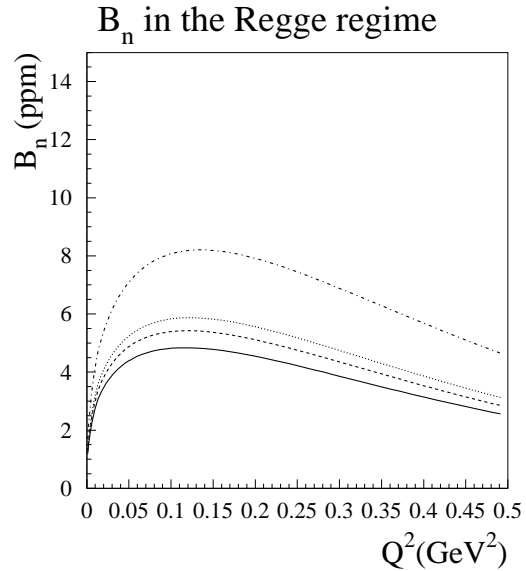}}
\caption{Results for $B_n$ as function of $Q^2$ for different values of the 
LAB energy of the electron: 6 GeV (solid line), 9 GeV (dashed line), 12 GeV 
(dotted line) and 45 GeV (dashed-dotted line).}
\label{ssafw}
\end{figure}
In Fig. \ref{ssafw}, beam normal spin asymmetry for the proton target is 
shown as function of $Q^2$ for four different values of $lab$ electron beam 
energy. In Fig. \ref{ssafw}, the sum of forward $A_3+A_6\sim\sigma_{tot}$ and 
backward $A_1\sim\pi\pi$-exchange is shown. However, the impact of the latter 
is negligibly small in the kinematics shown and is far below $1\%$ for all 
energies considered. The reason for that are the both suppression factors 
$m_{\pi,\rho}\over(s-M^2)$ and $Q_2\over(s-M^2)$ together with the Regge 
suppression due to the pion trajectory, as compared to the Pomeron. 
So, the $\log^2m$ enhancement of $B_n$ is absolutely irrelevant in this 
extreme forward kinematics. Recalling that this $\log^2m$ term factorizes the 
backward Compton amplitude contribution, this is by no means a surprise, 
since it is unphysical for bacward mechanisms to dominate forward observables.
This result suggests in turn that the QRCS approximation 
should not be expected to work in the forward kinematics. In fact, this effect 
was observed on the example of $\pi N$ intermediate state contribution in 
\cite{marcbarbara}. However, if going to 
backward angles, this contribution increases as $\sim |t|\ln^2(|t|/m^2)$ and, 
in general, the approximation is justified.
In the forward kinematics, the energy dependence in Eq.(\ref{eq:B_n_Ai})
originates thus mostly from the term $\ln{2E\over \sqrt{|t|}}$, 
so that the asymmetry increases logarithmically with the energy. 
This is an interesting observation and is on contrary with the 
dominant energy dependence of $B_n$ at low energies where it decreases with 
energy as $m/E$. It furthermore gives good outlook for measuring $B_n$ through 
high energy region at forward angles. Still, a full calculation (beyond the 
QRCS approximation) is needed to confirm this behaviour.
Comparing to the calculation of 
\cite{afanas_qrcs}, the quite different way the real Compton scattering amplitudes 
enter the final result for $B_n$ leads to inversed energy dependence and 
consequently, the largest disagreement between the two calculations amounts in 
an order of magnitude for the JLab energy, lowest among those considered. 
This discrepancy will be discussed in details in the next section.
\section{Discussion}
\label{sec:discussion}
As it has been noticed before, the presented calculation is in disagreement 
with the result of Ref. \cite{afanas_qrcs}. Their approach consisted in taking only 
the part of doubly virtual Compton tensor which survives in the exact forward 
limit of real Compton scattering (see Eq. (14) of \cite{afanas_qrcs}) 
\beqn
M^{\mu\nu}&=&\left\{-(PK)^2g^{\mu\nu}-(q_1q_2) P^\mu P^\nu\right.\\
&&\left.+(PK)\left[P^\mu q_1^\nu+P^\nu q_2^\mu\right]\right\} {\cal A}\nn,
\label{eq:tensor_rho2}
\eeqn
with ${\cal A}$ the forward amplitude of doubly virtual Compton scattering 
which is then related to the proton structure function $F_1$ and the total 
photoabsorption cross section $\sigma_{\gamma p}^{tot}$.
The forward spin independent doubly virtual Compton scattering tensor is 
usually written in terms of the DIS structure functions $W_1,W_2$,
\beqn
W^{\mu\nu}&=&O_1^{\mu\nu}W_1(\nu,Q^2)+O_2^{\mu\nu}W_2(\nu,Q^2)\nn\\
&=&\left(-g^{\mu\nu}+\frac{q_2^\mu q_1^\nu}{(q_1\cdot q_2)}\right) W_1
\label{eq:dis_str_f}\\
&+& 
\frac{1}{M\nu} \left(P^\mu+\frac{Pq_1}{(q_1\cdot q_2)}q_2^\mu\right) 
\left(P^\nu+\frac{Pq_2}{(q_1\cdot q_2)}q_1^\nu\right) W_2\nn,
\eeqn
where the tensors should be taken in the limit $q_1=q_2$. In the 
scaling limit, one has $W_1(\nu,Q^2)\to F_1(x)$, $W_2(\nu,Q^2)\to F_2(x)$, 
and the Callan-Gross relation implies $F_2=2xF_1$.
We rewrite now the tensor of Eq.(\ref{eq:tensor_rho2}) in terms of the DIS 
tensors $O_1^{\mu\nu},O_2^{\mu\nu}$:
\beqn
M^{\mu\nu}=(PK)^2
\left\{O_1^{\mu\nu}-\frac{q_1\cdot q_2}{P\cdot K}O_2^{\mu\nu}\right\}{\cal A}.
\eeqn
\indent
We proceed by projecting this tensor onto the basis of Prange, 
make use of ${q_1\cdot q_2\over K^2}=2+{Q_1^2+Q_2^2\over 2K^2}$, and obtain 
for the amplitudes' combination which multiplies the $\ln^2m$ term:
\beqn
B_1+B_2+\nu(B_2+B_4)&=&{(PK)^2\over2M}
{Q_1^2+Q_2^2\over 2K^2}{\cal A},
\label{a1_cross_sec}
\eeqn
\indent
The kinematical limit of 
Ref. \cite{afanas_qrcs} corresponds to neglecting $t$ comparing to the photons' 
virtualities. Then, $q_1=q_2=K$, and the right hand side is non-zero.
Instead, throughout this work, the non-forward RCS kinematics with 
the on-shell photons is used, and the contribution of the cross section 
(i.e., helicity conserving helicity amplitudes) to Eq.(\ref{a1_cross_sec}) 
vanishes. It is an 
interesting phenomenon, that the result of taking the limit 
$Q_{1,2}^2\to0$ and $t\to0$ is order depending.\\
\indent
The dependence on the order of taking the limit
$Q^2\to0$ and low photon energy limit has been observed before in the 
context of polarizabilities of RCS and generalized polarizabilities of VCS 
\cite{andreas}, and even for the leading term in photon energy of the 
forward RCS vs. forward doubly VCS amplitude $F(\nu,Q^2)$ \cite{reviewdr}:
\beqn
\lim_{\nu\to0}\lim_{Q^2\to0} F(\nu,Q^2)&=&-{\alpha_{em}\over M}e_N 
(\vec{\varepsilon}'^*\cdot\vec{\varepsilon}),\\
\lim_{Q^2\to0}\lim_{\nu\to0} F(\nu,Q^2)&=&
{\alpha_{em}\over M}\kappa_N(2e_N+\kappa_N) 
(\vec{\varepsilon}'^*\cdot\vec{\varepsilon}),\nn
\eeqn
with $e_N$ and $\kappa_N$ the charge and the a.m.m. of the nucleon 
($1$ and $1.79$ for the proton and $0$ and $-1.91$ for the neutron, 
respectively). 
So, it is not completely unexpected that the two different limits do not match.
\begin{figure}[h]
\vspace{-1cm}
{\includegraphics[height=8cm]{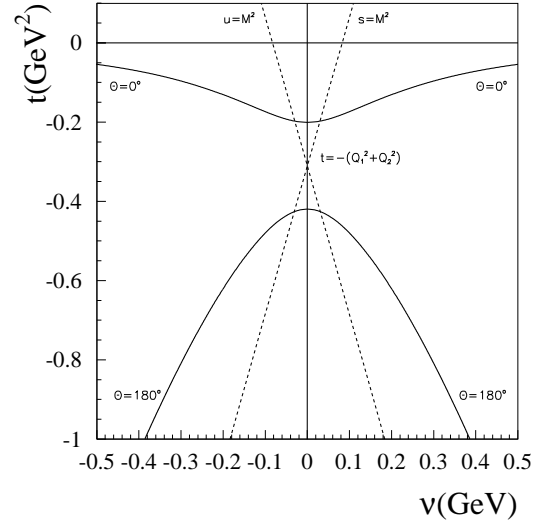}}
\caption{Mandelstam plot for doubly virtual Compton scattering with 
$Q_1^2=0.3$ GeV$^2$ and $Q_2^2=0.01$ GeV$^2$. For further details, see text.}
\label{vvcs_mandelstam}
\end{figure}
In Fig. \ref{vvcs_mandelstam}, the Mandelstam plot for the general case of 
doubly virtual Compton scattering is shown for $Q_1^2=0.3$ GeV$^2$ and 
$Q_2^2=0.01$ GeV$^2$. The plot shows different kinematical regions on the 
plane $\nu={Pq_1\over M}$ and $t$. The energy threshold for the reaction in 
the $s$-channel, $s=M^2$, 
is represented by the straight line $\nu={t+Q_1^2+Q_2^2\over4M}$, so that the 
$s$-channel reaction (direct Compton scattering) is possible to the right of 
this line. The requirement that the scattering angle takes physical values, 
restricts allowed values for the variables $\nu$ and $t$ to the area between 
the two curves depicted in the right half-plane. The upper line corresponds to 
$\cos\Theta=1$ (forward scattering), while the lower one ot $\cos\Theta=-1$ 
(backward scattering). The physical scattering can only occur inside this 
region. The forward line crosses the $t$ axis at
$t=-(\sqrt{Q_1^2}-\sqrt{Q_2^2})^2$, while the backward line at 
$t=-(\sqrt{Q_1^2}+\sqrt{Q_2^2})^2$.
The $u$-channel (crossed Compton reaction) kinematical regions are 
obtaine by $\nu\to-\nu$. The variable $\nu$ should not be confused with the 
energy of the virtual photon, which is sometimes denoted as $\nu$, as well. 
In the lab frame, the initial photon energy is given by 
$\omega_1={w^2-M^2+Q_1^2\over2M}$, while 
$\nu={2(w^2-M^2)+t+Q_1^2+Q_2^2\over4M}$, with $w^2=(p+q_1)^2$ the 
invariant mass squared of the intermediate hadronic state. For each value 
of the external variables $s$ and $t$ which one measures in a 
$ep$-scattering experiment, the integral over the intermediate electron phase 
space involves $\int_{M^2}^{(\sqrt{s}-m)^2}dw^2$ and $\int dQ_1^2dQ_2^2$. The 
latter integral is performed over the ellyptic areas shown in 
Fig. \ref{fig:kinbounds}, unambiguously defined by $s$, $t$ and $w^2$. 
In terms of the Mandelstam plot in Fig. \ref{vvcs_mandelstam}, 
this corresponds to integrating over a series of such Mandelstam plots for 
each pair of $Q_1^2,Q_2^2$ at the value of $t$ fixed by the experimental 
kinematics. 
In the special case of $Q_1^2=Q_2^2$ which is always contained within the 
integral, as can be seen in Fig. \ref{fig:kinbounds}, the Mandelstam plot is 
deformed, so that the forward line becomes $t=0$. If $Q_1^2\neq Q_2^2$, 
however, $t=0$ is only approached asymptotically for $\nu\to\infty$. 
This is the reason why the limit $t\to0$ is not well defined for the 
general case of the doubly virtual Compton scattering: this limit takes one 
outside of the physical region of the reaction. If one preforms the limit 
$Q_{1,2}^2\to0$ first, keeping $t\neq0$, one enssures that the line $t=0$ is 
contained within the physical region, and the limit can be performed.\\
\indent
Technically, the difference between the 
two limits originates from treating as small quantity either the momentum 
transfer (forward doubly VCS) or the virtualities of the photons (RCS). 
We note that the squared logarithm enhancement occurs if the hadronic 
amplitudes is non-zero in the exact limit of hard collinear photons. 
Therefore, the leading (in $\log^2m$) contribution should correspond to 
Compton scattering with the photon virtualities of the order 
$Q_{1}^2=-(k-k_1)^2\approx 2mE$ and the momentum transfer of the $ep$ reaction
which one measures on the experiment.
It means that in order to justify the limit of \cite{afanas_qrcs}, 
one has to go to external momentum transfers smaller than $2mE$ which is of 
the order of $10^{-3}$ GeV$^2$ in the kinematics considered here. Apart from 
difficulty in achieving this kinematics experimentally, it is suppressed by 
the kinematical 
factors in the general expression of Eq.(\ref{eq:bn_general})for $B_n$, which 
ensures this observable to vanish in the exact forward direction. Basing on 
these considerations, we see that in order to obtain adequate estimate of the 
double log effect for $B_n$, one has to split the full doubly virtual 
Compton amplitude into the 
non-forward real Compton amplitude which gives the desirable effect, plus the 
rest which is regular in the QRCS point and can at most give a $\ln m$ effect. 
\section{Summary}
\label{sec:summary}
In summary, we presented a calculation of the beam normal spin asymmetry in 
the kinematical regime of high energies ($E_{lab}=6-45$ GeV) and low 
momentum transfer to the target, $Q^2\leq0.5$ (GeV/c)$^2$. This observable 
obtains its leading contribution from the imaginary part of the 
two photon exchange graph times the Born amplitude and is 
directly related to the imaginary part of doubly virtual Compton scattering
The resulting loop integral obtains a large contribution from the kinematics 
when both exchanged photons are nearly real and collinear to the external 
electrons. We adopt the QRCS or equivalent photons approximation which allows 
to take the hadronic part in the external kinematics out of the integral and 
to perform the integration over the electron phase space analytically. 
For the hadronic part, we use the full real Compton scattering amplitude 
and show, that both forward (no proton and photon helicity flip) 
and backward (photon helicity flipping) amplitudes do contribute in this 
observable in the forward kinematics. For the forward Compton amplitude, we 
use optical theorem as the input. For backward Compton amplitude, we provide 
an explicite calculation which is due to two pion exchange in the $t$-channel. 
The resulting values of the asymmetry are in the range $4-8$ ppm for the 
energies in the range $6-45$ GeV. We find furthermore that the double 
logarithmic enhancement does not dominate $B_n$ in forward regime since it 
comes with helicity-flip Compton amplitude which highly suppresses this 
behaviour. Finally, we make an interesting observation that the 
non-analyticity of the doubly virtual Compton scattering 
leads to the order dependence of taking limits $t\to0$ and $Q_{1,2}^2\to0$, 
which is a generalization of the known problem of matching the low energy 
limits of real and doubly virtual Compton scattering.
\begin{acknowledgments} The author is grateful to Prof. M.M.Giannini for 
continuous support and to Dr. M.J. Ramsey-Musolf and Dr. V. Cirigliano for 
numerous discussions and for thorough reading of the manuscript. 
The work was supported by Italian MIUR and INFN, and by US Department of 
Energy Contract DE-FG02-05ER41361.
\end{acknowledgments}
\section{Appendix A: Integrals over electron phase space}
In this section we present calculation of the integrals over electron phase 
space appearing in the QRCS approximation. First we calculate the scalar 
integral $I_0$,
\beqn
I_0\;=\;\int_0^{k_{thr}}\frac{k_1^2dk_1}{2E_1(2\pi^3)}
\int\frac{d\Omega_{k_1}}{(k-k_1)^2(k'-k_1)^2},
\eeqn
where the upper integration limit $k_{thr}$ corresponds to the inelastic 
threshold (i.e. pion production), 
$k_{thr}=\sqrt{\frac{(s-(M+m_pi)^2)^2}{4s}-m^2}$. We 
next introduce integration over the Feynman parameter 
${1\over ab}=\int_0^1{dx\over [a+(b-a)x]^2}$. We chose the polar axis such as 
$\vec{k}_1\cdot(\vec{k}-x\vec{q})=k_1|\vec{k}-x\vec{q}|\cos\Theta_1$ with 
$|\vec{k}-x\vec{q}|^2=k^2+x(1-x)t$ and perform angular integration.
We furthermore change integration over $dk_1$ 
to integration over dimensionsless $z=E_1/E$
\beqn
I_0&=&
\frac{1}{-8\pi^2t}\int_{m\over E}^{{E_{thr}}\over E}
\frac{dz}{\sqrt{z^2-\frac{m^2}{E^2}-\frac{4m^2}{t}(1-z)^2}}\\
&&\;\;\;\;\;\;\cdot
\ln\frac{\sqrt{z^2-\frac{m^2}{E^2}-\frac{4m^2}{t}(1-z)^2}+\sqrt{z^2-\frac{m^2}{E^2}}}
{\sqrt{z^2-\frac{m^2}{E^2}-\frac{4m^2}{t}(1-z)^2}-\sqrt{z^2-\frac{m^2}{E^2}}},
\nn
\eeqn
\indent
To perform the integration 
over the electron energy, we follow here the main details of the calculation 
in Appendix A of Ref.\cite{afanas_qrcs}. The result reads
\beqn
I_0\;=\;\frac{1}{-32\pi^2}
\left\{
\ln^2\left(\frac{-t}{m^2}\frac{E_{thr}^2}{E^2}\right)\,+\,
8Sp\left({{E_{thr}}\over E}\right)\right\},
\eeqn
where $Sp(x)$ is the Spence or dilog function, 
$Sp(x)=-\int_0^1{dt\over t}\ln(1-xt)$ with $Sp(1)=\pi^2/6$. In the high 
energy limit, ${{E_{thr}}\over E}\to1$, we recover the result of 
Ref.\cite{afanas_qrcs}.\\
\indent
We next turn to the vector integral
\beqn
I_1^\mu&=&\int_0^{k_{thr}}\frac{k_1^2dk_1}{2E_1(2\pi^3)}
\int d\Omega_{k_1}\frac{k_1^\mu}{(k-k_1)^2(k'-k_1)^2}\nn\\
&=&I_{1P}P^\mu\,+\,I_{1K}K^\mu\,.
\eeqn
\indent
It cannot depend on $q^\mu$ due to the symmetry of $I_1$ under 
interchanging $k$ and $k'$. To determine these two coefficients we have a 
system of equations,
\beqn
K^0I_{1K}\,+\,P^0I_{1P}&=&I_1^0\;=\;
{1\over16\pi^3}\int\frac{d^3\vec{k}_1}{Q_1^2Q_2^2}\\
-tI_{1K}\,+\,4PKI_{1P}&=&4K_\mu I_1^\mu\;=\;
\int\frac{d^3\vec{k}_1}{(2\pi^3)E_1Q_1^2}\nn\\
&\equiv&I_2.\nn
\eeqn
Using the same approach as for $I_0$, we obtain for $I_1^0$:
\beqn
I_1^0&=&\frac{E}{-8\pi^2t}\int_{m\over E}^{{E_{thr}}\over E}
\frac{zdz}{\sqrt{z^2-\frac{m^2}{E^2}-\frac{4m^2}{t}(1-z)^2}}\\
&&\;\;\;\;\;\;\;\;\cdot
\ln\frac{\sqrt{z^2-\frac{m^2}{E^2}-\frac{4m^2}{t}(1-z)^2}+\sqrt{z^2-\frac{m^2}{E^2}}}
{\sqrt{z^2-\frac{m^2}{E^2}-\frac{4m^2}{t}(1-z)^2}-\sqrt{z^2-\frac{m^2}{E^2}}}
\nn\\
&=&\frac{E}{-4\pi^2t}
\left\{{{E_{thr}}\over E}\ln{\sqrt{-t}\over m}{{E_{thr}}\over E}\right.\\
&&\;\;\;\;\;\;\;\;\;\;\;\;\left.\,+\,
\left(1-{{E_{thr}}\over E}\right)\ln\left(1-{{E_{thr}}\over E}\right)
\right\}.
\eeqn
\indent
Finally, we consider the integral $I_2$ where we perform angular integration,
\beqn
I_2&=&{1\over(2\pi^3)}\int\frac{d^3\vec{k}_1}{E_1Q_1^2}\\
&=&
\pi\int_{m\over E}^{{E_{thr}}\over E}dz
\ln\frac{z-\frac{m^2}{E^2}+\sqrt{1-\frac{m^2}{E^2}}\sqrt{z^2-\frac{m^2}{E^2}}}
{z-\frac{m^2}{E^2}-\sqrt{1-\frac{m^2}{E^2}}\sqrt{z^2-\frac{m^2}{E^2}}}.\nn
\eeqn
\indent
After integrating by parts and changing variables consequently 
$z={m\over E}\cosh y$ and $y=\ln t$, we arrive to
\beqn
I_2&=&2\pi{{E_{thr}}\over E}\ln{{2{E_{thr}}\over m}}\\
&+&2\pi\left(1-{{E_{thr}}\over E}\right)\ln\left(1-{{E_{thr}}\over E}\right).
\nn
\eeqn
\indent
In these last two integrals, it is important to keep ${{E_{thr}}\over E}$ 
unequal to 1 till the end to ensure the convergence of the integral. 
Solving the system of linear equations for the coefficients, we obtain:
\beqn
I_{1P}&=&\pi\frac{s-M^2}{M^4-su}{{E_{thr}}\over E}\ln{2E\over Q}\nn\\
I_{1K}&=&\frac{1}{Q^2}I_2-\frac{4PK}{Q^2}I_{1P}.
\eeqn
\section{Appendix B: Scalar integrals for helicity flip amplitude}
The vector and tensor integrals can be reduced 
to the scalar ones by means of standard methods \cite{passarino_veltman}. 
The remaining integrals to be calculated are the two, three, and four-point 
scalar integrals. Here we are only interested in the imaginary part of these, 
therefore there are only three integrals with non-zero imaginary part:
the two-point integral
\beqn
C^\pi_0&=&{\rm Im}\int\frac{d^4p_\pi}{(2\pi)^4}
\frac{1}{p_\pi^2-m_\pi^2}\frac{1}{(P+K-p_\pi)^2-M^2}\nn\\
&=&\frac{1}{8\pi}\frac{|\vec{p}_\pi|}{\sqrt{s}},
\eeqn
with $|\vec{p}_\pi|=\sqrt{\frac{(s-M^2+m_\pi^2)^2}{4s}-m_\pi^2}$;
the three-point one
\beqn
B^\pi_0&=&{\rm Im}\int\frac{d^4p_\pi}{(2\pi)^4}
\frac{1}{p_\pi^2-m_\pi^2}\frac{1}{(k-p_\pi)^2-m_\pi^2}\nn\\
&&\;\;\;\;\;\;\;\;\cdot
\frac{1}{(P+K-p_\pi)^2-M^2}\nn\\
&=&-\frac{1}{8\pi(s-M^2)}\ln\frac{2E_\pi}{m_\pi},
\eeqn
and finally, the four-point integral:
\beqn
A^\pi_0&=&{\rm Im}\int\frac{d^4p_\pi}{(2\pi)^4}
\frac{1}{(k-p_\pi)^2-m_\pi^2}\frac{1}{p_\pi^2-m_\pi^2}\nn\\
&&\;\;\;\;\;\;\;\;\cdot
\frac{1}{(k'-p_\pi)^2-m_\pi^2}\frac{1}{(P+K-p_\pi)^2-M^2}\nn\\
&=&\frac{1}{8\pi Q^2(s-M^2+m_\pi^2)}
\nn\\
&\cdot&
\frac{1}{\sqrt{1+{{4m_\pi^2E^2}\over Q^2p_\pi^2}}}
\ln\frac{\sqrt{1+{{4m_\pi^2E^2}\over Q^2p_\pi^2}}+1}
{\sqrt{1+{{4m_\pi^2E^2}\over Q^2p_\pi^2}}-1}.
\eeqn
\indent
These integrals should however be reggeised as described in Section 
\ref{sec:results} by substituting the Regge propagator instead of the Feynman 
one. Denoting 
$t_1=(k-p_\pi)^2$ and $t_2=(k'-p_\pi)^2$ the momentum transferred by the pions 
in the $t$-channel, we have for the reggeized version of scalar integrals:
\beqn
(C^\pi_0)^R&=&\frac{1}{32\pi^2}\frac{p_\pi}{\sqrt{s}}\int d\Omega_\pi
(t_1-m_\pi^2){\cal P}_\pi^R(\alpha_\pi(t_1))\nn\\
&&\;\;\;\;\;\;\;\;\;\;\;\;\;\;\;\;\;\;
\cdot(t_2-m_\pi^2){\cal P}_\pi^R(\alpha_\pi(t_2))
\eeqn
\beqn
(B^\pi_0)^R&=&\frac{1}{32\pi^2}\frac{p_\pi}{\sqrt{s}}\int d\Omega_\pi
(t_1-m_\pi^2)\\
&&\;\;\;\;\;\;\;\;\;\;\;\;\;\;{\cal P}_\pi^R(\alpha_\pi(t_1))
{\cal P}_\pi^R(\alpha_\pi(t_2)),\nn
\eeqn
\beqn
\!\!\!\!\!\!(A^\pi_0)^R
\,=\,\frac{1}{32\pi^2}\frac{p_\pi}{\sqrt{s}}\!\!\int \!\!d\Omega_\pi
{\cal P}_\pi^R(\alpha_\pi(t_1))
{\cal P}_\pi^R(\alpha_\pi(t_2))
\eeqn
\indent
Similarly, in the case of the $\rho$-exchange in the $s$-channel, the integrals
 with non-zero imaginary part are:
\beqn
C^\rho_0&=&{\rm Im}\int\frac{d^4p_\rho}{(2\pi)^4}
\frac{1}{p_\rho^2-m_\rho^2}\frac{1}{(P+K-p_\rho)^2-M^2}\nn\\
&=&\frac{1}{8\pi}\frac{|\vec{p}_\rho|}{\sqrt{s}},\nn
\eeqn
\beqn
B^\rho_0&=&{\rm Im}\int\frac{d^4p_\rho}{(2\pi)^4}
\frac{1}{p_\rho^2-m_\rho^2}\frac{1}{(k-p_\rho)^2-m_\pi^2}\nn\\
&&\;\;\;\;\;\;\;\;\;\;\;\;\;\;\cdot\frac{1}{(P+K-p_\rho)^2-M^2}\nn\\
&=&-\frac{1}{8\pi(s-M^2)}\ln\frac{2E_\rho(s-M^2)}{Mm_\rho^2},\nn
\eeqn
\beqn
A^\rho_0&=&{\rm Im}\int\frac{d^4p_\rho}{(2\pi)^4}
\frac{1}{(k-p_\rho)^2-m_\pi^2}\frac{1}{p_\rho^2-m_\rho^2}\nn\\
&&\;\;\;\;\;\;\;\;\;\;\;\;\;\;\cdot
\frac{1}{(k'-p_\rho)^2-m_\pi^2}\frac{1}{(P+K-p_\rho)^2-M^2}\nn\\
&=&\frac{1}{8\pi Q^2(s-M^2+m_\rho^2)}
\frac{1}{\sqrt{1+{{4\sigma^2}\over Q^2p_\rho^2}}}\nn\\
&\cdot&
\ln\frac{\sqrt{1+{{4\sigma^2}\over Q^2p_\rho^2}}+1}
{\sqrt{1+{{4\sigma^2}\over Q^2p_\rho^2}}-1},
\eeqn
with $|\vec{p}_\rho|=\sqrt{\frac{(s-M^2+m_\rho^2)^2}{4s}-m_\rho^2}$, and 
$\sigma^2=E^2m_\rho^2-EE_\rho(m_\rho^2-m_\pi^2)+{1\over4}(m_\rho^2-m_\pi^2)^2$.
If neglecting the pion mass, $\sigma={{Mm_\rho^2}\over 2s}$.
The reggeization procedure is the same as for $\pi N$- intermediate state.

\end{document}